\documentclass{moriond}

\usepackage{amsmath, amssymb}
\usepackage{tensor}
\usepackage{physics}

\newcommand{\X}{Y}

\begin{document}
\vspace*{4cm}
\title{A new look on black hole perturbations in modified gravity}

\author{Hugo Roussille}
\address{Universit\'e Paris Cit\'e,  CNRS, Astroparticule et Cosmologie, F-75013 Paris, France}

\maketitle\abstracts{We study the linear perturbations about a nonrotating black hole solution of Horndeski's theory, using a systematic approach that extracts the asymptotic behaviour of perturbations (at spatial infinity and near the horizon) directly from the first-order radial differential system governing these perturbations instead of finding Schr\"odinger-like equations for their dynamics. We illustrate this method in the case of a specific black hole solution. The knowledge of the asymptotic behaviours of the perturbations paves the way for a numerical computation of the quasinormal modes.  Finally, the asymptotic form of the modes also signals   some pathologies in the scalar sector of the solution considered here.}

\section{Introduction}

The dawn of gravitational wave (GW) astronomy has spurred a renewed interest in possible deviations from General Relativity (GR), which could be detected for example in the ringdown phase of a binary black hole merger. This phase is described by linear perturbations about a background stationary black hole solution, and the waves emitted correspond to a superposition of quasinormal modes, whose frequencies are quantised \cite{Kokkotas:1999bd,Konoplya:2011qq}. One expects that modified gravity models would predict QNMs that differ from their  GR counterpart and the detailed analysis of the GW signal, commonly called ``black hole spectroscopy", represents an invaluable window to test General Relativity and to look for specific signatures of modified gravity \cite{Berti:2018vdi}. So far, QNMs  have been investigated only for a few models of modified gravity \cite{Berti:2018vdi}.

In these proceedings, we present a new  approach  for the study  of  black holes perturbations illustrated  in the context of Horndeski scalar-tensor theories. We start by describing the computation of QNMs in the case of General Relativity, then explain how modified gravity theories lead to new challenges for their computation. We finally present an algorithm that allows us to determine the asymptotic behaviour of black hole perturbations without casting them into a second-order Schr\"odinger equation. This algorithm paves the way for the computation of QNMs spectra in modified gravity theories.

\section{Black hole perturbations in General Relativity}
\label{sec:BH-pert-GR}

In this section, we review  the standard procedure to derive the equations of motion  for  the perturbations of a  Schwarzschild black hole in general relativity, originally obtained  by  Regge and Wheeler \cite{Regge:1957td} for the axial, or odd-parity, modes and Zerilli \cite{Zerilli:1970se} for the polar, or even-parity, modes. These equations  can be shown to reduce to a Schr\"odinger-like equation with an effective potential characterising the ``dynamics'' of the linear perturbations.

\subsection{Regge-Wheeler-Zerilli gauge}

Let us consider a Schwarzschild background  metric $\tensor{\overline{g}}{_\mu_\nu}$. This metric is written as
\begin{equation}
	\tensor{\overline{g}}{_\mu_\nu} \dd{x^\mu} \dd{x^\nu} = -A(r) \dd{t}^2 + \frac{1}{B(r)} \dd{r}^2 + r^2 \dd{\Omega^2} \,,
	\label{eq:spherical-metric}
\end{equation}
with $A(r) = B(r) = 1 - \mu/r$, $\mu$ being a constant of integration. We study metric perturbations by introducing the perturbed metric $\tensor{g}{_\mu_\nu}=\tensor{\overline{g}}{_\mu_\nu}+ \tensor{h}{_\mu_\nu}$, where the $\tensor{h}{_\mu_\nu}$ denote the linear  perturbations of the metric.

One can separate the components of $h_{\mu\nu}$ into two classes, depending on their behaviour with respect to parity transformations: axial perturbations correspond to odd parity while polar perturbations correspond to even parity. In the following, we focus solely on polar perturbations. These perturbations are decomposed onto spherical harmonics and Fourier transformed; once the remaining gauge freedom is fixed, one can prove that the perturbed Einstein's equations can be written as a two-dimensional first order system:
\begin{eqnarray}
\label{eq:systeme-2-eqs}
\dv{\X}{r} = M_\text{polar}(r) \X \,,
\end{eqnarray}
where the two components of $\X$ are some combination of the components of $h_{\mu\nu}$.

\subsection{Schrödinger equation and potential}

The system of Eq.~\eqref{eq:systeme-2-eqs} can be cast into a single  second order Schr\"odinger-like equation for a unique dynamical variable \cite{Regge:1957td,Zerilli:1970se}. To do this, we consider the general (linear) change of vector $\X(r) = P(r) \hat \X (r)$, where $\hat \X$ is a new column vector and the two dimensional invertible matrix $P$ has  not been fixed at this stage. We also define a new radial coordinate $r_*$ and introduce the ``Jacobian'' of the transformation $n(r) \equiv \dv*{r}{r_*}$.

It can be shown that it is possible to find a matrix $P$ such that the new system satisfied by $\hat{\X}$ takes the canonical form
\begin{equation}
	\label{eq:systeme-reduit}
	\dv{\hat \X}{r_*}  = \begin{pmatrix} 0 & 1 \\ V_\text{polar}(r) - \omega^2 & 0\end{pmatrix} \hat \X \, ,
\end{equation}
where the potential $V(r)$ depends on  $r$, but not on $\omega$. This form can be interpreted as a wave propagation equation for $\hat{\X}_1$:
\begin{equation}
\dv[2]{\hat{\X}_1}{r_*}  + \left(\omega^2 - V_\text{polar}(r)\right) \hat{\X}_1 = 0 \,.
\label{eq:schrodinger-like-general}
\end{equation}
This wave equation will be extremely useful for the computation of QNMs, as we show below.

\subsection{Quasinormal modes}

Finding quasi-normal modes requires to impose the appropriate boundary conditions: the modes must be outgoing at infinity and ingoing at the horizon. One can show that $V_\text{polar}$ goes to zero at infinity and at the horizon, which implies that Eq.~\eqref{eq:schrodinger-like-general} becomes asymptotically $\hat{\X}_1''(r_*) + \omega^2 \hat{\X}_1(r_*) \approx 0$, where $ \approx$ is an equality up to sub-leading corrections. Therefore, at both boundaries, the function $\hat{\X}_1$ behaves like 
\begin{equation}
\hat{\X}_1(r) \approx  {\cal A} \, e^{+ i\omega r_*} + {\cal B} \, e^{-i\omega r_*}  \,,
\label{eq:behav-infinity-schwarzschild}
\end{equation}
where  ${\cal A}$ and ${\cal B} $ are integration constants which take different values at the horizon and at infinity.

We can interpret each term as a radially propagating wave by putting back the $e^{-i \omega t}$ factor: the terms proportional to ${\cal A}_\text{hor}$ and ${\cal A}_\infty$ are outgoing while the terms proportional to ${\cal B} _\text{hor}$ and ${\cal B} _\infty$ are ingoing. Imposing ${\cal A} _\text{hor} = 0$ and ${\cal B} _\infty = 0$ severely restricts  the possible values of $\omega$: the authorized values are the quasinormal modes of the Schwarzschild black hole. These values can be found numerically by integrating the Schr\"odinger-like equation \cite{Konoplya:2011qq}.

\section{New challenges in modified gravity}

\subsection{The impact of scalar perturbations}

The perturbations of a black hole in a scalar-tensor theory of gravity are similar to the perturbations of the Schwarzschild solution described in Sec.~\ref{sec:BH-pert-GR}, with a new degree of freedom added since the scalar field can be perturbed too. This scalar field perturbation is also decomposed onto spherical harmonics, and after Fourier transform it is parametrised by only one function $\delta\phi(r)$. As scalar perturbations must be of even parity, the system of equations governing polar perturbations must contain two degrees of freedom in the case of scalar-tensor theories. As a consequence, the first-order system presented for Schwarzschild in Eq.~\eqref{eq:systeme-2-eqs} becomes 4-dimensional:
\begin{equation}
	\dv{\X}{r} = M_\text{p+s} \X \,,
	\label{eq:systeme-4d-polar}
\end{equation}
where $M_\text{p+s}$ contains the dynamics of both the polar gravitational and the scalar modes.

Naturally, one initially tries to find a way to cast such a system onto two decoupled Schr\"odinger equations, similarly to the Schwarzschild case presented in Eq.~\eqref{eq:systeme-reduit}. This requires finding $P$ such that
\begin{equation}
	\X = P \hat{\X} \qq{and} \dv{\hat{\X}}{r_*} = \begin{pmatrix}
		0 & 1 & 0 & 0 \\
		V_\text{p}(r) - \frac{\omega^2}{c_\text{p}^2} & 0 & 0 & 0 \\
		0 & 0 & 0 & 1 \\
		0 & 0 & V_\text{s}(r) - \frac{\omega^2}{c_\text{s}^2} & 0
	\end{pmatrix} \hat{\X} \,.
\end{equation}

In general, finding such a transformation is impossible as the system of equations is complicated. The aim of the work presented in these proceedings is to find the asymptotic behaviour of $\X$ at the black hole horizon and at infinity, in order to rule out solutions that exhibit pathologies and allowing one to get the necessary boundary conditions for the computation of QNMs, without obtaining the wave propagation equations in their Schr\"odinger form. In order to do this, we must expand the first order system~\eqref{eq:systeme-4d-polar} at the horizon and at infinity.

\subsection{Asymptotic behaviour of the BCL black hole}

In order to illustrate our method, we study a solution \cite{Babichev:2017guv} obtained for a subset of  Horndeski theories. The metric sector of this solution corresponds to the general form of Eq.~\eqref{eq:spherical-metric} with functions $A$ and $B$ different from the Schwarzschild case, and a nonzero scalar field $\phi$. We call this solution ``BCL black hole'' from the names of the authors.

In the case of the BCL black hole, the expression of the first-order matrix $M_\text{p+s}$ at infinity is
\begin{equation}
	M_\text{p+s} = M_2 r^2 + M_1 r + M_0 + \dots \,,
\end{equation}
where $M_2$ has zeros everywhere except for $(M_2)_{2,1} = \omega^2$, and $M_1$ and $M_0$ are more complex constant matrices. It is not possible to recover the asymptotic behaviour for $\X$ from this expansion. Indeed, if one truncates the matrix at order 1, one obtains
\begin{equation}
	\dv{\X}{r} = M_2 r^2 \X \qq{so} \X = \exp(M_2 r^3/ 3) \X_c = \qty(I_4 + \frac{r^3}{3} M_2) \X_c\,,
\end{equation}
with $\X_c$ a constant and $I_4$ the 4-dimensional identity matrix. This means that $\X$ does not behave in a wave-like fashion at infinity: the expected behaviour would be the one of Eq.~\eqref{eq:behav-infinity-schwarzschild}.

In fact, finding the asymptotic behaviour of a system by truncating it is only possible if the system is diagonal. In the case of the BCL black hole, this is obviously not the case. The solution is to use a mathematical algorithm \cite{balserComputationFormalFundamental1999,Langlois:2021xzq}, that gives a systematic way to compute the asymptotic behaviour of $\X$ for any kind of system.

The algorithm provides us with a matrix $P$ such that $\X = P \tilde{\X}$, and the first-order system for $\tilde{\X}$ has a matrix $\tilde{M}_\text{p+s}$ such that
\begin{equation}
	\tilde{M}_\text{p+s} = \text{diag}(-i\omega, i\omega, -\sqrt{2}\omega, \sqrt{2}\omega) + \mathcal{O}(1/r) \,.
\end{equation}
This implies that the behaviour of $\tilde{\X}$ at infinity is a sum of four modes: two gravitational modes $\mathfrak{g}^{+\infty}_\pm$ and two scalar modes $\mathfrak{s}^{+\infty}_\pm$, with
\begin{equation}
	\mathfrak{g}^{+\infty}_\pm(r) \,\approx\, e^{\pm i \omega r} \qq{and} \mathfrak{s}^{+\infty}_\pm(r) \,\approx\, e^{\pm \sqrt{2} \omega r} \,.
\end{equation}
The identification of the former modes as gravitational comes from the comparison with the Scwharzschild case.

One can then conclude that the scalar sector does not propagate towards infinity for this black hole solution: this pathology, while not a proof of instability, shows us that the solution is not relevant physically. One can then impose that $\tilde{\X}$ behave as $\mathfrak{g}^{+}$ at infinity and recover the behaviour of $\X$; doing the same computation at the horizon leads to two boundary conditions on $\X$, which is enough to compute the quasinormal modes of the BCL spacetime \cite{Langlois:2021aji}.

\section{Conclusion}

In these proceedings, we have studied linear black hole perturbations in the context of a specific black hole solution to Horndeski theories. The method used in GR relying on the reformulation of the dynamics in terms of a Schr\"odinger-like equation is not applicable here; instead of using it, we developped a very generic method that enables one to obtain the asymptotic behaviours of the perturbations at spatial infinity and near the black hole horizon from the first-order system extracted directly from the perturbed Einstein's equations. The knowledge of these asymptotic behaviours  is essential to define and compute the quasi-normal modes,  characterised by outgoing conditions at spatial infinity and ingoing conditions at the horizon.

\section*{References}

\bibliography{biblio_roussille_proceedings}

\end{document}